\documentstyle{mn}

\input psfig
\newif\ifAMStwofonts

\def\etal{et al.~}
\def\Msun{{\rm\,M_\odot}}
\def\Lsun{{\rm\,L_\odot}}
\def\kms{{\rm\,km\,s^{-1}}}

\def\kmskpc{{\rm\,km\,s^{-1}\,kpc^{-1}}}
\def\deg{^{\circ}}

\def\spose#1{\hbox to 0pt{#1\hss}}
\def\lta{\mathrel{\spose{\lower 3pt\hbox{$\sim$}}
    \raise 2.0pt\hbox{$<$}}}
\def\gta{\mathrel{\spose{\lower 3pt\hbox{$\sim$}}
    \raise 2.0pt\hbox{$>$}}}
\newdimen\hssize
\newdimen\hdsize
\ifoldfss
  \ifCUPmtlplainloaded \else
    \NewTextAlphabet{textbfit} {cmbxti10} {}
    \NewTextAlphabet{textbfss} {cmssbx10} {}
    \NewMathAlphabet{mathbfit} {cmbxti10} {} 
    \NewMathAlphabet{mathbfss} {cmssbx10} {} 
  \fi
  \ifAMStwofonts
    \ifCUPmtlplainloaded \else
      \NewSymbolFont{upmath} {eurm10}
      \NewSymbolFont{AMSa} {msam10}
      \NewMathSymbol{\upi}     {0}{upmath}{19}
      \NewMathSymbol{\umu}     {0}{upmath}{16}
      \NewMathSymbol{\upartial}{0}{upmath}{40}
      \NewMathSymbol{\leqslant}{3}{AMSa}{36}
      \NewMathSymbol{\geqslant}{3}{AMSa}{3E}

    \fi
  \fi
\fi 

\ifnfssone
  \newmathalphabet{\mathit}
  \addtoversion{normal}{\mathit}{cmr}{m}{it}
  \addtoversion{bold}{\mathit}{cmr}{bx}{it}
  \newmathalphabet{\mathbfit} 
  \addtoversion{normal}{\mathbfit}{cmr}{bx}{it}
  \addtoversion{bold}{\mathbfit}{cmr}{bx}{it}
  \newmathalphabet{\mathbfss} 
  \addtoversion{normal}{\mathbfss}{cmss}{bx}{n}
  \addtoversion{bold}{\mathbfss}{cmss}{bx}{n}
  \ifAMStwofonts
    \ifCUPmtlplainloaded \else
      \UseAMStwoboldmath
      \makeatletter
      \new@mathgroup\upmath@group
      \define@mathgroup\mv@normal\upmath@group{eur}{m}{n}
      \define@mathgroup\mv@bold\upmath@group{eur}{b}{n}
      \edef\UPM{\hexnumber\upmath@group}
      \new@mathgroup\amsa@group
      \define@mathgroup\mv@normal\amsa@group{msa}{m}{n}
      \define@mathgroup\mv@bold\amsa@group{msa}{m}{n}
      \edef\AMSa{\hexnumber\amsa@group}
      \makeatother
      \mathchardef\upi="0\UPM19
      \mathchardef\umu="0\UPM16
      \mathchardef\upartial="0\UPM40
      \mathchardef\leqslant="3\AMSa36
      \mathchardef\geqslant="3\AMSa3E
    \fi
  \fi
\fi 

\ifnfsstwo
  \DeclareMathAlphabet{\mathbfit}{OT1}{cmr}{bx}{it}
  \SetMathAlphabet\mathbfit{bold}{OT1}{cmr}{bx}{it}
  \DeclareMathAlphabet{\mathbfss}{OT1}{cmss}{bx}{n}
  \SetMathAlphabet\mathbfss{bold}{OT1}{cmss}{bx}{n}
  \ifAMStwofonts
    \ifCUPmtlplainloaded \else
      \DeclareSymbolFont{UPM}{U}{eur}{m}{n}
      \SetSymbolFont{UPM}{bold}{U}{eur}{b}{n}
      \DeclareSymbolFont{AMSa}{U}{msa}{m}{n}
      \DeclareMathSymbol{\upi}{0}{UPM}{"19}
      \DeclareMathSymbol{\umu}{0}{UPM}{"16}
      \DeclareMathSymbol{\upartial}{0}{UPM}{"40}
      \DeclareMathSymbol{\leqslant}{3}{AMSa}{"36}
      \DeclareMathSymbol{\geqslant}{3}{AMSa}{"3E}
    \fi
  \fi
\fi 

\ifCUPmtlplainloaded \else
  \ifAMStwofonts \else 
    \def\upi{\pi}
    \def\umu{\mu}
    \def\upartial{\partial}
  \fi
\fi

\title[Bar driven evolution of NGC 4570]
{Bar driven evolution of S0s: the edge-on galaxy NGC~4570}
\author[F. C. van den Bosch and E. Emsellem]
       {Frank C. van den Bosch$^{1,2}$ and Eric Emsellem$^{3,4}$\\
    $^1$ Department of Astronomy, University of Washington, Box
            351580, Seattle, WA 98195-1580, USA \\
    $^2$ Hubble Fellow \\
    $^3$ European Southern Observatory, Karl-Schwarzschild
            Strasse 2, 85748 Garching b. M\"unchen, Germany \\
    $^4$ Centre de Recherche Astronomique de Lyon, Observatoire de Lyon,
            9 av. Charles Andr\'e, 69561 Saint-Genis Laval Cedex, France}

\date{Accepted . Received}
      
\pagerange{\pageref{firstpage}--\pageref{lastpage}}
\pubyear{1998}

\begin{document}

\maketitle

\label{firstpage}

\begin{abstract}
  We present circumstantial evidence that the central region of the
  edge-on S0 galaxy NGC~4570, which harbors a 150~pc scale nuclear
  disc in addition to its main outer disc, has been shaped under the
  influence of a small ($\sim 500$~pc) bar.  This is based on the
  finding of two edge-on rings, whose locations are consistent with
  the Inner Lindblad and Ultra Harmonic Resonances of a rapidly
  tumbling triaxial potential. Observed features in the photometry and
  rotation curve nicely correspond with the positions of the main
  resonances, strengthening the case for a tumbling bar potential. The
  relative blue colour of the ILR ring, and the complete absence of
  any detected ISM, indicates that the nuclear ring is made of
  relatively young ($\lta 2$ Gyr) stars. We discuss a possible
  secular-evolution scenario for this complex multi-component galaxy,
  which may also apply to many other S0-galaxies with observed rings
  and/or multiple disc-components.
\end{abstract}


\section{Introduction}

High resolution images obtained with the Hubble Space Telescope (HST)
have revealed the presence of bright stellar discs with very small
scale lengths of $\sim 20$~pc in a number of S0 galaxies (van den
Bosch \etal 1994; Lauer \etal 1995; van den Bosch, Jaffe \& van der
Marel 1998).  These nuclear discs are photometrically and
kinematically decoupled from the main (outer) discs which have an
inner cut-off radius; the so-called Freeman type II discs (Freeman
1970). Such a multi-component structure suggests a complex formation
scenario. The main question is whether this multi-component structure
was imposed during the formation of the galaxy or whether it resulted
from secular evolution.

Baggett, Baggett \& Anderson (1996) argued that the inner-truncated
discs could be linked with the presence of a bar. Since the nuclear
regions of these S0s exhibit high line strength values (Fisher, Franx
\& Illingworth 1995), it has been suggested that they formed from
radial inflow of pre-enriched gas. Small scale (nuclear) bars have
been invoked as an efficient way to transport gas inwards and
accumulate mass in the central regions of galaxies (Friedli \&
Martinet 1993; Wada \& Habe 1995).  Bar-induced secular evolution
could thus be a possible mechanism by which to form the observed
nuclear discs. In this scenario the inner cut-off of the outer main
disc would occur at the co-rotation radius of the bar, which drives
gas into the central region where the nuclear disc forms.
Unfortunately, there is a conspiracy in that nuclear discs can only be
easily detected in nearly edge-on galaxies, whereas the photometric
bar signatures (e.g., isophote twists) are confined to the more
face-on systems.

In this paper we present the case of NGC~4570, an edge-on S0 galaxy
with a nuclear stellar disc. Based on the study of the photometric and
kinematical substructures in this galaxy (Section~2), we present evidence
for the presence of a single pattern speed associated with a small bar
(Section~3). This evidence is based mainly on the locations of two
edge-on stellar rings. In Section~4 we discuss a secular evolution
scenario for NGC~4570 under the influence of this bar potential.
Conclusions are drawn in Section~5.

\section{Substructures in NGC 4570}
\label{sec:sub}

All data discussed here, except the ground-based image, have been
presented by van den Bosch, Jaffe \& van der Marel (1998, hereafter
BJM98). We refer the interested reader to this article for further
details on the data and its reduction.

\subsection{Photometry}
\label{sec:photo}

BJM98 obtained $U$, $V$, and $I$ band images of NGC~4570 using the
Wide Field and Planetary Camera 2 (WFPC2) aboard the HST. In order to
build a detailed photometric model of NGC~4570 (see
Section~\ref{sec:model}) we use an additional ground-based $V$ band
image (resolution of $\sim 1\farcs6$ FWHM) which provides a larger
field of view. This image was obtained with the 2m telescope of
Pic-du-Midi Observatory by Jean-Luc Nieto and kindly provided to us by
Patrick Poulain. All images were reduced in the classical way, and
normalized in $\Lsun\,{\rm pc}^{-2}$ (in the corresponding Cousin
bands). The ground-based and HST $V$-band images show excellent
agreement with each other after the difference in resolution is
properly taken into account.

\subsubsection{Unsharp masking}
\label{sec:mask}

\begin{figure}
\psfig{figure=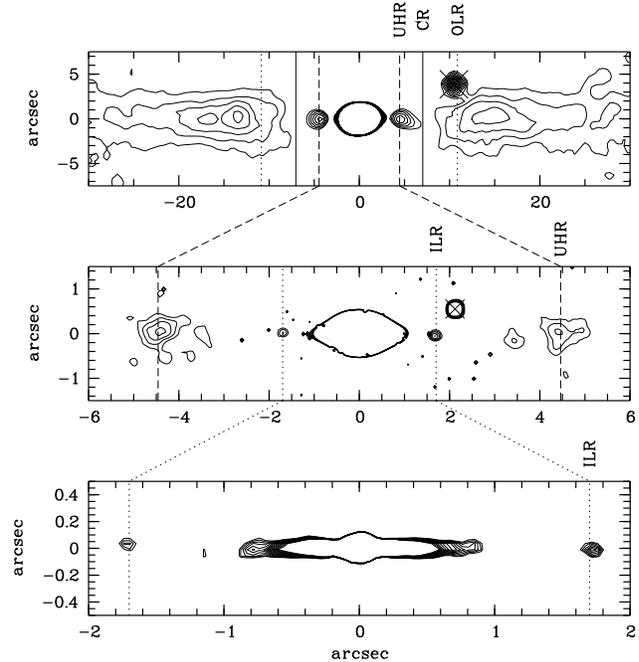,width=\hssize}
\caption{Unsharp masking versions of the ground-based $V$-band image 
  (top panel) and of the WFPC2 image (two bottom panels) of NGC~4570.
  The upper panel reveals the presence of the outer disc with an inner
  cut-off at $\sim 7\arcsec$, whereas the lower panel clearly shows
  the presence of the nuclear disc inside $1\farcs0$. In addition to
  this double-disc structure, the different panels reveal two
  symmetric peaks at $\pm 1\farcs69$ and $\pm 4\farcs44$ from the
  centre along the major-axis. The outer peaks appear on both the
  ground-based image and the HST image.  The central part of each
  frame has a different appearance due to the differences in the size
  of the filtering window.  Two star-like objects in the fields have
  been marked by a cross to avoid confusion. The vertical lines
  correspond to the presumed resonances (ILR, UHR, CR and OLR as
  indicated; see Section~\ref{sec:res}).}
\label{fig:rings}
\end{figure}

In order to enhance the substructures in our images we have performed
a simple unsharp masking of our WFPC2 and ground-based images
(Fig.~\ref{fig:rings}): the HST image clearly reveals the thin nuclear
disc inside the central arcsecond (cf. BJM98), and the ground-based
image shows that the outer disc is really a Freeman type-II disc, with
an inner cut-off radius at $\sim 7\arcsec$. There are thus two nested
disc-like components.
 
In addition to the double disc structure, the unsharp masked images
reveal two sets of symmetric peaks, $K_{\pm 1}$ and $K_{\pm 2}$,
located at respectively $\pm 1\farcs69$ and $\pm 4\farcs44$ from the
centre along the major-axis (Fig.~\ref{fig:rings}). $K_{\pm 1}$ are
very sharp and marginally resolved, whereas $K_{\pm 2}$ are diffuse
(FWHM $\sim 1\arcsec$) and less contrasted against the background
galaxy. The symmetry of these peaks suggests that they correspond to
two individual rings located in the equatorial plane of the galaxy and
seen nearly edge-on. In the following, we refer to these structures as
the nuclear or $K_1$ ring, and the outer or $K_2$ ring.

%
%
%

\subsubsection{Isophote shapes}

\begin{figure}
\psfig{figure=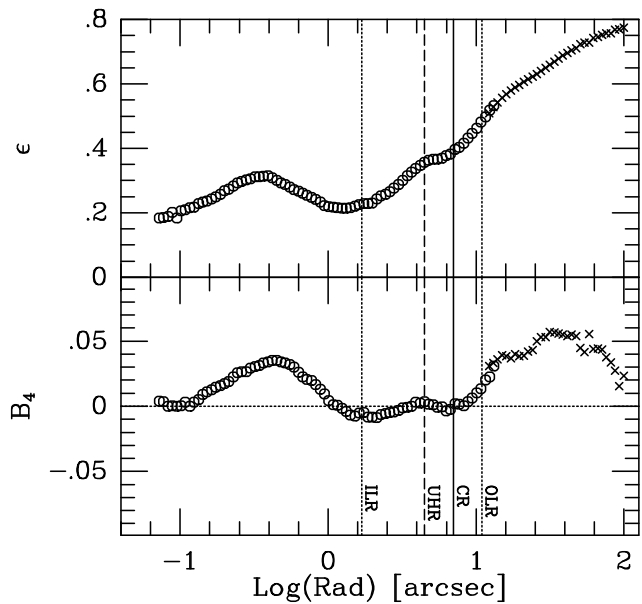,width=\hssize}
\caption{Ellipticity in $V$ (top panel), and the $B_4$ coefficient (bottom 
  panel) along the major-axis of NGC~4570, derived from the WFPC2
  images (circles) and the ground-based image (crosses). The
  locations of the presumed ILR, UHR, CR and OLR (see
  Section~\ref{sec:res}) are indicated by the vertical lines.}
\label{fig:ell}
\end{figure}

In Fig.~\ref{fig:ell} we plot the ellipticity ($\epsilon$) and the
$B_4$ coefficient as a function of the semi major-axis radius (for
both the HST and ground-based images). The $B_4$ coefficient describes
deviations of isophotes from pure ellipses, such that positive $B_4$
indicates discy isophotes and negative $B_4$ corresponds to
isophotes that are more boxy than an ellipse (see Jedrzejewski 1987).
The different components of NGC~4570 clearly show up in these
diagrams:
\begin{itemize}
\item the maximum of $B_4$ and $\epsilon$ at $\sim 0\farcs4$
  corresponds to the highest contrast between the nuclear disc and the
  central spheroid.
\item the position of the nuclear ring corresponds to a small local
  maximum in both the ellipticity ($\epsilon \sim 0.22$) and $B_4$
  ($\sim -0.005$) profiles. However this is entirely due to the
  presence of the $K_{\pm 1}$ peaks along the major-axis. When their
  contribution is removed, the location of $K_1$ corresponds to the
  absolute minima in the ellipticity and the $B_4$ profiles (i.e.,
  minimum flattening and maximum boxiness).
\item from the end of the nuclear disc outwards, the ellipticity
  profile rises up to 0.37 at the radius of the $K_2$ ring, is flat in
  between $4\farcs5$ and $7\arcsec$, and rises again outwards.
\item the outer ring corresponds to a local maximum of the $B_4$
  coefficient.  Again, this may be entirely due to the presence of the
  $K_2$ peaks.  Finally, the $B_4$ profile shows a local minimum at
  the radius of $7\arcsec$.
\end{itemize}

An investigation of the detailed isophotal parameters of NGC~4570 thus
reveals three characteristic radii: $\sim 1\farcs69$, inside of which
the nuclear disc dominates the projected surface brightness; $\sim
4\farcs5$, where a local bump in the ellipticity and $B_4$ profiles is
visible; and $7\arcsec$, outside of which the contribution of the
outer disc to the surface brightness starts to dominate. In between
$\sim 1\farcs69$ and $7\arcsec$ the slightly boxy spheroid ($B_4 \sim
-0.01$) dominates the projected surface brightness.

%
%
%
%

\subsubsection{Colours}
\label{sec:col}

\begin{figure}
\psfig{figure=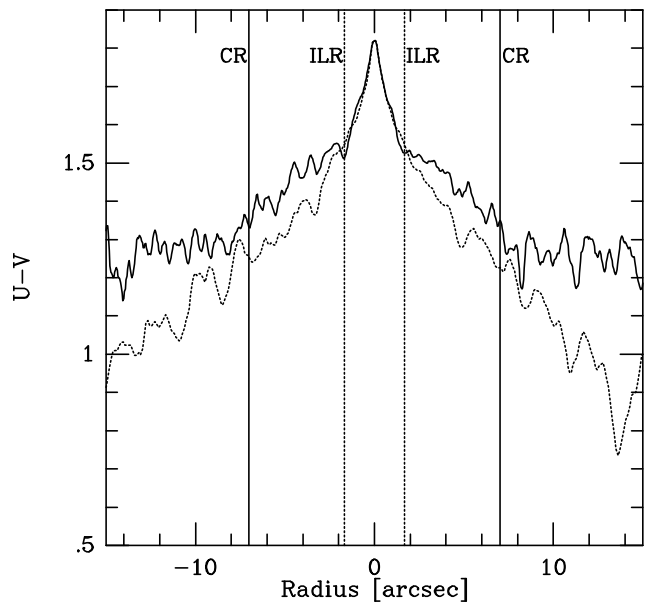,width=\hssize}
\caption{The $U - V$ colour profiles, averaged over a $0\farcs3$ slit along
  the major (solid curve) and minor (dotted curve) axes of NGC~4570.
  The abscissa of the minor-axis profile has been scaled with the
  inverse of the local flattening measured in the $V$-band. The
  locations of the presumed ILR and CR (see Section~\ref{sec:res}) are
  indicated by the vertical lines.}
\label{fig:majcol}
\end{figure}

The three separate regions discussed above also show up in the colour
images. Fig.~\ref{fig:majcol} shows the $U-V$ profiles averaged over
a slit of $0\farcs3$ along the major and minor axis respectively. The
colour image, derived from the WFPC2 images of NGC~4570, has been
slightly smoothed with an adaptive filter which damps the noise but
preserves the characteristic features. Along the major axis,
the radius of the nuclear ring and $R = 7\arcsec$
clearly mark the changes in the slope of the colour
gradient, and thus delimit three distinct regions:
\begin{itemize}
\item Outside $\sim 7\arcsec$, the cut-off radius of the outer disc,
  the colour profile is nearly flat with $U - V \sim 1.25$.
\item In between $K_1$ and $R \sim 7\arcsec$ the $U - V$ colour slowly
  increases inwards.
\item The region inside the nuclear ring exhibits a very steep
  increase in $U-V$.
\end{itemize}
Along the minor axis, where the light is bulge-dominated, the $U-V$
profile exhibits a steep, continuous gradient. The outer disc, which
seems to have a constant $U-V$ colour, is clearly much redder than the
outer parts of the bulge at the corresponding radius (scaled with the
local ellipticity, see Fig.~\ref{fig:majcol}). Fisher \etal (1996)
have shown that this is a common feature amongst S0s, and that the
line-strength profiles show a similar behavior, except for ${\rm
  H}\beta$. This suggests that the (main) disc is of similar age as
the bulge, but of higher metallicity.
%
%
%
%

\subsection{Kinematics}

\begin{figure}
\psfig{figure=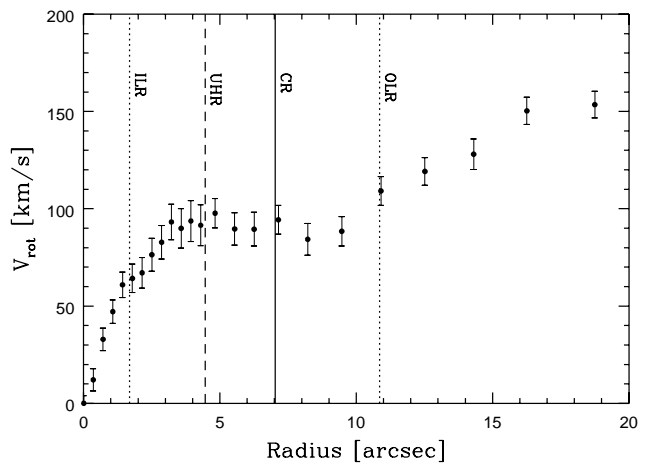,width=\hssize}
\caption{Rotation velocity along the major-axis of NGC~4570, 
  derived from the WHT red spectra. The locations of the presumed ILR,
  UHR, CR and OLR (see Section~\ref{sec:res}) are indicated by the
  vertical lines.}
\label{fig:velo}
\end{figure}

In addition to the photometry, BJM98 used the 4.2m William Herschel
Telescope (WHT) at La Palma to obtain long-slit spectra of NGC~4570
at three different position angles: major axis, minor axis and an
offset axis parallel to the major axis. The spectra have a spectral
resolution of $\sigma_{\rm instr} = 9\kms$, and were obtained with a
slit-width of $1\farcs0$ and seeing FWHM of $1\farcs1$.  After
standard reduction of the spectra, the rotation velocities ($V_{\rm
  rot}$) and rms velocities ($V_{\rm rms}$), which are the first and
second order moments of the line-of-sight velocity profile (hereafter
VP), were derived. The characteristic radii mentioned above mark
again specific features in the observed rotation curve along the major
axis of NGC~4570 (Fig.~\ref{fig:velo}):
\begin{itemize}
\item Inside the nuclear ring, the stellar component exhibits a nearly
  rigid body rotation with $V_{\rm rot} \sim 390 \kmskpc$.  This value
  obviously depends on the spatial resolution.
\item At the radius of $K_1$ (which marks the end of the contribution
  of the nuclear disc), the gradient of the rotation curve suddenly
  drops and the rotation velocity increases out to $\sim 4\arcsec$.
\item In between $4\arcsec$ and $\sim 10\arcsec$ the rotation velocity 
  is roughly constant but starts to rise again further outwards where 
  the outer disc dominates the projected surface brightness.
\end{itemize}
%
%
%
%

\section{Evidence for a single pattern speed}

\subsection{The rings}
\label{sec:rings}

In Section~\ref{sec:mask}, we have shown evidence for the presence of
two rings, at radii of $1\farcs69$ and $4\farcs46$ respectively.  As
emphasized by Buta \& Combes (1996), {\em rings are always associated
  with gas}. Gas rings are a commonly observed phenomenon in barred
galaxies (Buta 1995), where they originate from the torques provided
by the tumbling non-axisymmetric bar structure (Schwarz 1984). Such
bar-induced rings are often observed at the so-called Lindblad
Resonances: the Inner Lindblad Resonance or ILR ($\Omega_p =\Omega -
\kappa / 2$), and the Outer Lindblad Resonance or OLR ($\Omega_p
=\Omega + \kappa / 2$).  Here $\Omega_p$ is the pattern speed of the
bar, and $\Omega$ and $\kappa$ are respectively the circular and the
radial epicyclic frequencies in the equatorial plane of the galaxy.
Rings also often form at the 4/1 Ultra Harmonic Resonance (UHR) which
corresponds to $\Omega_p = \Omega - \kappa / 4$ (Schwarz 1984; see
also Jungwiert \& Palou\v s 1996 for a more general treatment).  It is
important to note that gas rings can also be formed by viscous
shearing at the turnover radius of the velocity curve (Lesch \etal
1990). However, Jungwiert \& Palou\v s (1996) argued that inelastic
collisions significantly reduce the turbulent motions thus rendering
this mechanism inefficient. It could anyway not explain the presence
of {\it two} rings in NGC~4570.

In order to examine whether the positions of the observed $K_{\pm 1}$
and $K_{\pm 2}$ features are consistent with resonance radii
associated with a tumbling bar potential, we determine the potential
of NGC~4570. We construct a model for the spatial luminosity
distribution which, after projection and convolution with the point
spread function (PSF), fits the observed surface brightness
distribution.  The absolute scale of the potential is set by the
mass-to-light ratio of the model, which we determine from solving the
Jeans equations (see Binney \& Tremaine 1987) and fitting to the
observed kinematics.

\subsection{The photometric model}
\label{sec:model}

\begin{figure}
\psfig{figure=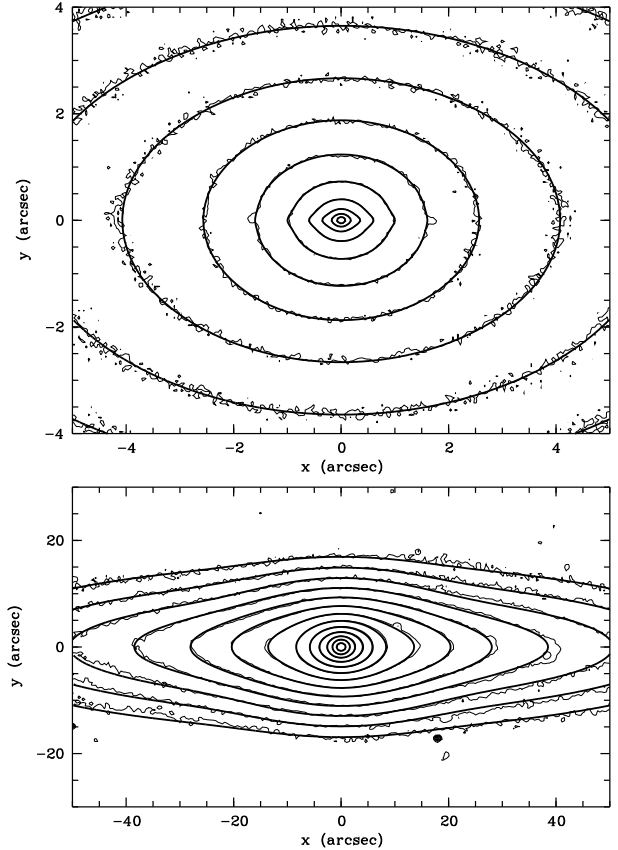,width=\hssize}
\caption{$V$ band isophotes (thin lines) from the WFPC2 data 
  (FWHM $\sim 0\farcs1$, top panel) and from the ground-based data
  (FWHM $\sim 1\farcs6$, bottom panel). The MGE analytical fit (thick
  lines) has been superimposed in both panels taking into account the
  different resolutions. The slight discrepancies with the
  ground-based data are due to a slight asymmetry in the isophotes of
  NGC~4570 at large scales.  The brightest isophotes are 14 and 16
  mag. arcsecs$^{-2}$ in the top and bottom panels respectively; the
  step is 0.5 mag.arcsec$^{-2}$ in both panels.}
\label{fig:isophot}
\end{figure}

We fit both $V$-band images using the Multi-Gaussian Expansion
(MGE) technique including the effect of the respective PSFs (Emsellem,
Monnet \& Bacon 1994). This technique is particularly well suited for
galaxies with complex multi-component structures such as NGC~4570.
The resulting model provides an excellent fit to the available
photometric data from the central HST pixel ($\sim 0\farcs045$) up to
a radius of $\sim 105\arcsec$ along the major-axis
(Fig.~\ref{fig:isophot}). The residual between the WFPC2 image and the
MGE model reveals the $K_{\pm 1}$ and $K_{\pm 2}$ peaks already
uncovered via the unsharp masking (the $K_{\pm 1}$ peaks are directly
visible in the top panel of Fig.~\ref{fig:isophot}).The thinness of
the nuclear disc indicates that the inclination angle $i > 82\deg$
(Scorza \& van den Bosch 1998).  Throughout we assume an inclination
angle of $i = 90\deg$ (i.e,.  edge-on). The results presented here are
only very weakly dependent on this assumption. Using this inclination
angle we deproject the unconvolved MGE model to obtain an analytical
form of the three dimensional luminosity distribution.
%
%
%
%

\subsection{The dynamical model}
\label{sec:dyn}

We solved the Jeans equations for the MGE model of NGC~4570 assuming a
constant mass-to-light ratio and an isotropic dispersion tensor. The
mass-to-light ratio was adjusted such as to obtain the best fit to the
rms velocities, which are independent of the velocity anisotropy
$\sigma_{\phi}/\sigma_{R}$ of the model (here $\sigma$ is the spatial
velocity dispersion). Assuming a distance of 23 Mpc for NGC~4570
($1\arcsec \sim 112$~pc), we find $M/L_{V} = 3.51$. These Jeans
models, which are based on the oversimplifying assumption that the
phase-space distribution function only depends on the two classical
integrals of motion (energy $E$ and vertical angular momentum $L_z$),
do not fit the details of the observed kinematics. However, this is of
no importance here, since we are merely interested in obtaining an
estimate of the mass-to-light ratio.

\subsection{The pattern speed}
\label{sec:res}

\begin{figure}
\psfig{figure=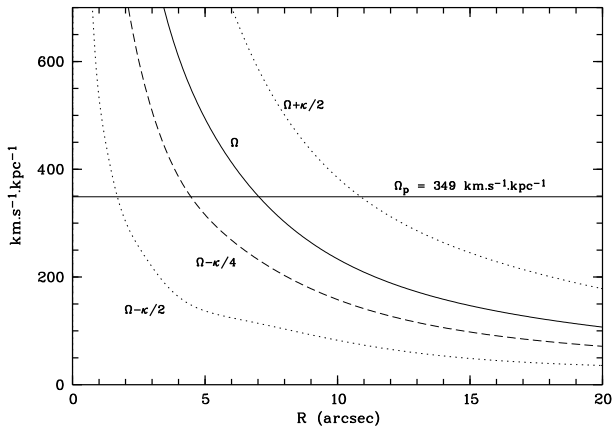,width=\hssize}
\caption{Resonance diagram for NGC~4570: $\Omega$ and 
  $\kappa$ are the circular and the epicycle frequencies. The
  horizontal line corresponds to a pattern speed of $349 \kmskpc$.}
\label{fig:resonance}
\end{figure}

Using the MGE model of NGC~4570 and the mass-to-light ratio derived
from the Jeans modeling, we calculate the potential and its
corresponding circular and epicyclic frequencies as functions of the
cylindrical radius $R$.  We thus derive the resonance diagram of
NGC~4570, which presents the main radial orbital frequencies versus
the radius $R$ in the equatorial plane (Fig.~\ref{fig:resonance}).
Given the small distance of the $K_1$ ring from the centre, we
assume it to be at the ILR ($R = 1\farcs69$).  According to the
resonance diagram, this implies a pattern speed of $349 \kmskpc$ and
sets the locations of the UHR, co-rotation (CR, $\Omega_p = \Omega$),
and OLR at $4\farcs46$, $7\farcs03$ and $10\farcs86$, respectively.
The locations of these resonances are indicated as vertical lines in
Fig.~\ref{fig:rings} to~\ref{fig:velo}.

These radii are in excellent agreement with the ones emphasized in
Sect.~\ref{sec:sub}.  The radius of the UHR ($4\farcs46$) corresponds
nearly exactly with the observed location of the secondary $K_{\pm 2}$
peaks ($\pm 4\farcs44$).  Both the ILR and the UHR are located in the
region dominated by bulge light. The nuclear disc is located inside
the ILR. The outer disc starts just outside CR ($R \sim 7\arcsec$) and
dominates the light outside the OLR.  
%
%
%
%

The UHR ring, when present, usually encircles the bar which would
therefore have a radial extent of $\sim 500$~pc. Both the high pattern
speed and the small size of the bar suggest that the bar is comparable
to nuclear bars observed in several (close to face-on) spiral
galaxies. M\"ollenhoff, Matthias \& Gerhard (1995) presented the case
of M94, a ringed Sab galaxy, where they detect the presence of a
small bar (major axis of $\sim 0.7$ kpc) with a pattern speed of $290
\kmskpc$ and which contributes only $\sim 14$\% of the total light
inside its extent. The bar-like system of M~94 may be very similar to
what we observe in NGC~4570.  

\subsection{Robustness of the resonances radii}
\label{sec:robust}

The circular and epicycle frequencies are defined in an axisymmetric
potential and hence only used here as approximations. It is striking
to see how well the observed locations of the peaks are consistent
with a simple resonance diagram derived from such axisymmetric model.
In strong bars, ILR rings are often elongated (typical axis ratios
lower than 0.8) and perpendicular to the bar: they follow the main
family of $x_2$ orbits which passes into the $x_1$ family (elongated
parallel to the bar) upon crossing the ILR (e.g., Contopoulos \&
Grosbol 1989). In such case, the locations of the peaks associated
with these elongated rings depend on the viewing angle and in general
will not coincide exactly with the resonance radii derived as outlined
above. This tells us that the potential of NGC~4570 may not depart
very strongly from axisymmetry, i.e., the bar potential is weak (see
also Section~\ref{sec:kinem}).

We tested the dependence of the location of the resonances on the
model in three ways. We first derived the resonance diagram for a new
MGE model whose central part was fitted using the $I$ band HST image:
the UHR, CR and OLR resonances are then moved to $4\farcs22$,
$6\farcs68$ and $10\farcs19$ respectively, which corresponds to a
decrease of about 5\%.  Secondly, we allowed the $M/L_V$ of the
different components to vary, in order to take into account a possible
variation of $M/L_V$ with $(R, z)$: the amplitude of the $M/L$
variation was chosen consistently with the observed colour gradients.
The derived empirical errors for the location of the individual
resonances (UHR, CR and OLR) were all smaller than 10\%. We finally
checked the effect of the uncertainty of the inclination angle, and
found it to be negligible ($\Delta R <1$\% for $i > 82\deg$).  The
predicted radii for the UHR, CR, and OLR resonances are thus found to
be rather robust. In what follows, we use the former values derived
from an edge-on model with constant mass-to-light ratio ($M/L_V =
3.51$).

\subsection{Nature and origin of the rings}
\label{sec:nature}

As mentioned in Section~\ref{sec:rings} gas rings are a natural
consequence of the presence of a bar.  However, we did not detect any
emission lines in our WHT spectra, which should include the H$\beta$
and [OIII] emission lines.  Furthermore, NGC~4570 has neither been
detected in X rays, nor in HI or 100 $\mu$m flux (Knapp \etal 1989;
Wrobel 1991). At HST resolution the $K_{\pm 1}$ peaks represent about
10\% of the local surface brightness in the $V$ band (which covers the
wavelengths of the [OIII] emission lines). In the $I$ band, in which
no strong emission lines are expected, the $K_{\pm 1}$ peaks still
contain about 7\% of the local light. This demonstrates that the light
associated with $K_{\pm 1}$ is mostly stellar. The same holds for the
UHR-ring.

According to our MGE model, NGC~4570 has a total luminous mass of
$\sim 5.8 \times 10^{10}$~M$_{\odot}$. For a stellar population with a
mean age of $5\times 10^{9}$~yr, we estimate that the entire galaxy
produces $\sim 0.8 \Msun\,{\rm yr}^{-1}$ of gas through stellar mass
loss (Ciotti \etal 1991). Only the gas located inside CR is
transported inwards, whereas outside CR gas is transported towards the
OLR.  Approximately 20\% of the total mass of NGC~4570 is located
inside CR.  We therefore find that the mass of the ILR ring could be
generated in less than 40 Myr by stellar mass loss. This shows that
the absence of any detected ISM does not oppose strong difficulties
against interpreting the observed peaks as bar-induced resonance
rings, since the gas mass required is easily supplied by internal mass
loss.
 
\begin{figure}
\psfig{figure=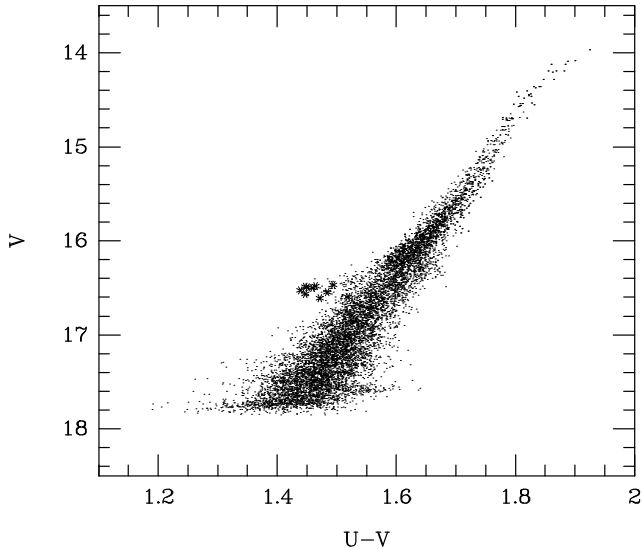,width=\hssize}
\caption{Colour-magnitude diagram of the central region of 
  NGC~4570. Each dot represents one pixel from the PC CCD. The
  asterisks indicate the pixels at the position of the two $K_{\pm 1}$
  peaks.}
\label{fig:colour}
\end{figure}

A colour-magnitude diagram of the inner region of NGC~4570 is
presented in Fig.~\ref{fig:colour}: here each dot represents a pixel
of the Planetary Camera (PC) CCD. The overall inclined shape of the
diagram reflects the colour gradient in the galaxy.  The asterisks
mark the pixels at the position of the two $K_{\pm 1}$ peaks.  They
are clearly bluer than their immediate surrounding.  Taking into
account the relative contribution of the $K_{\pm 1}$ peaks to the
local surface brightness (absolute maxima of 22\%, 13\% and 10.5\% in
the $U$, $V$ and $I$ bands respectively) we estimate the true colour
of the nuclear ring to be $U - V = 0.98 \pm 0.05$ and $V - I = 1.06
\pm 0.05$. According to the single burst, stellar population models of
Worthey (1994), these colours correspond to a rather young population
with an age of less than $\sim 2$ Gyr and a metallicity close to
Solar. Although stellar population models may be inaccurate (see e.g.,
Bruzual 1996), we suggest that the nuclear ring formed from gas which,
driven by a tumbling triaxial distortion, accumulated at the ILR and
underwent subsequent star formation in the last couple of Gyrs.

%
%
%
The case of the UHR ring is less clear as it is less contrasted
against the background light and appears as a rather clumpy structure.
It is slightly more contrasted in the $V$ band than in the $I$ band.
However we do not detect the $K_{\pm 2}$ peaks in the $U$ band.  If
the UHR ring were as blue as the ILR ring, the $K_{\pm 2}$ peaks
should represent about 10\% of the local luminosity in the $U$ band.
This is unfortunately too close to the noise level in the $U$ band
image for us to conclude on the true colour of the UHR ring.

It is also important to emphasize the fact that the colour gradient
shows a clear break at the CR radius, which marks the inner
cut-off of the outer disc. This is strikingly similar, at least
qualitatively, to the results obtained in the bar simulations by
Friedli, Benz \& Kennicutt (1994, see their Fig.~1, panel~d; see also
Friedli \& Benz 1995).  Their study shows that the presence of a bar
can significantly affect the abundance distribution: in their
simulations, the abundance gradient after 1~Gyr is nearly flat in the
disc outside CR, shallow in the bar, and very steep in the central
region where a nuclear starburst was triggered by an increased gas
fueling. In the case of NGC~4570, it would thus be important to
examine the link between the observed colours and e.g., the metal line
indices.

\subsection{Kinematic signatures of a bar}
\label{sec:kinem}

Although bars are difficult to detect photometrically in edge-on
systems, Kuijken \& Merrifield (1995) have shown that bars do produce
a strong kinematic signature. Due to the lack of available closed
orbits near co-rotation, the projected $(R,v_{\rm los})$-plane shows a
``figure-of-eight''-shape: at radii inside CR, the velocity profiles
are double peaked resulting in a characteristic figure-of-eight
variation with radius. The two peaks reflect orbits in the bar and in
the disc outside CR projected along the line-of-sight.  Since gas is
forced to move on closed, non-intersecting orbits, this kinematic
bar-signature shows up most clearly with emission lines.  However, as
long as not too many stars are on chaotic orbits, it should in
principle show up from absorption line spectra as well.  Indeed,
Kuijken \& Merrifield (1995) have found this characteristic
``figure-of-eight''-shape in the absorption line spectra of two
edge-on systems with peanut-shaped bulges.

\begin{figure}
\psfig{figure=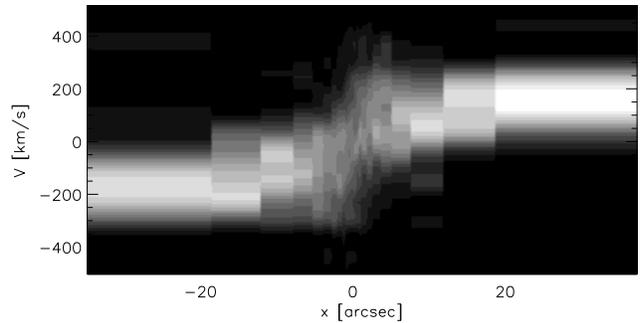,width=\hssize}
\caption{The $(R,v_{\rm los})$-diagram along the major axis of NGC~4570. 
  No bar-characteristic ``figure-of-eight''-shape is visible.}
\label{fig:lv}
\end{figure}

We have constructed a $(R,v_{\rm los})$-diagram for NGC 4570 from the
major axis WHT spectrum using the unresolved Gaussian decomposition
(UGD) algorithm developed by Kuijken \& Merrifield (1993). This method
models the VPs as the sum of a set of unresolved Gaussian
distributions with fixed means and dispersions.  The (non-negative)
weights of each of these Gaussians are then determined using quadratic
programming techniques. We spatially rebinned our spectrum to obtain a
signal-to-noise of at least 40 per $10 \kms$. Each VP was then modeled
as the sum of Gaussians with $44 \kms$ dispersion, and whose means are
separated by $66 \kms$ (we have checked that the choice of these
values does not influence the conclusion reached here).  The resulting
$(R,v_{\rm los})$-diagram is shown in Fig.~\ref{fig:lv}.  Although
some LOSVDs in the central $5\arcsec$ are ``flat-topped'' or
double-peaked, we do not detect a ``figure-of-eight'' in the
$(R,v_{\rm los})$-diagram.  Unfortunately, this does not provide a
definite test regarding the presence of a small scale bar in NGC~4570:
the bar could be weak and/or seen at an intermediate angle between
edge-on and end-on, both of which can impede detection of the
``figure-of-eight'' shape (Merrifield 1996).  Furthermore, since our
presumed bar is very small ($\sim 500$ pc), the
``figure-of-eight''-shape should occur inside that radius. However,
the (hot) bulge component dominates the light in that region, making
it very hard to detect the characteristics of the bar against this
bulge background.
%
%
%
%

\section{A secular evolution model}

It is generally believed that bars are formed through the process of
swing-amplification (Toomre 1981). However, the initial presence of an
ILR would prohibit the waves from reaching the centre and therefore
damp them. The steep stellar cusp in NGC~4570 ensures the presence of
such an ILR. So either the perturbation of the disc was strong enough
to saturate the ILR (e.g., due to a tidal interaction), or the cusp
(and therewith the ILR) was not present when the bar formed, but is
merely a product of the gas flow towards the centre induced by the
bar.  The nucleus ($R < 0\farcs5$) of NGC~4570 exhibits a large
H$\beta$ line strength value of $2.51\pm 0.48$ as well as a high Mg$_2
\sim 0.386 \pm 0.015$ (see BJM98). This, in addition to the central
red colours ($V-I \sim 1.37$ and $U-V \sim 1.8$), suggests the
presence of a relatively young and metal-rich stellar component.
Furthermore, the steep colour gradient starts inside the presumed ILR
(see Sect.~\ref{sec:col}). This strongly favors the hypothesis of the
nuclear disc and cusp being the result of pre-enriched gas accretion
with a subsequent burst of star formation. We therefore suggest that
the triaxial distortion was stronger in the past, and has been
partially dissolved due to the growth of a central mass-concentration
(and therewith the creation of a strong ILR).  Several studies have
shown that the growth of a central mass concentration weakens the bar,
and eventually dissolves it if the nucleus reaches a few percent
(typically $\sim 5\%$) of the total mass of the galaxy (e.g.,
Pfenniger \& Norman 1990; Norman, Sellwood \& Hasan 1996).  If we
assume that both the central cusp ($3.3 \times 10^8 \Msun$), and the
nuclear disc ($1.2 \times 10^8 \Msun$) were formed from bar-induced
gas inflow, the central region of NGC~4570 raised its mass by less
than $\sim 1\%$ of the total luminous mass ($5.8 \times 10^{10}
\Msun$).  Such a moderate increase of the central mass concentration
may not completely destroy the bar, but merely weaken it (Norman \etal
1996).  Hence a rapidly tumbling bar could survive in the central
region.  In this scenario, it is to be expected that when the bar
formed, no ILR was present. We found that upon subtracting the central
cusp from our MGE-model, the resulting potential has an $\Omega -
\kappa/2$-curve which is zero at $R=0$, and reaches a maximum of $\sim
330 \kmskpc$ at $R \sim 1\farcs0$. Therefore, a perturbation with a
pattern speed of $\Omega_p = 349\kmskpc$ in this potential will not
have an ILR, and can form a bar by means of swing amplification.

These are two further arguments that seem to suggest that the ILR ring
was formed after the bar. The first is the fact that the blue colour
of the nuclear ring indicates that it is very probably younger than
the stellar component inside the ILR. The second argument is based on
how well the rings can survive (partial) bar dissolution.  When the
bar dissolves, the potential shape evolves from triaxial to
axisymmetric. Whether the rings survive this evolution first depends
on the timescale of the dissolution of the bar.  If that timescale is
sufficiently smaller than the rotational period at the radius of the
ring $T_{\rm ring}$, stars in the ring experience violent relaxation
(Lynden-Bell 1967): stars loose energy if they are located along the
major-axis of the bar, and gain energy along the minor axis of the
bar: the ring is consequently destroyed. However, if the bar
dissolution timescale is much larger than $T_{\rm ring}$ the potential
change is adiabatic, and the actions of the orbits that make up the
ring are adiabatic invariants. The ring then slowly shapes itself
(i.e., becomes rounder) to the time-varying potential.  Sellwood
(1996) performed simulations, in which he mimicked the gas flow
towards the centre by gradually increasing the central density. He
found that the $m = 2$ harmonic (which sustains the bar) drops by a
factor 10 in less than one bar rotation period which is $\sim 2 \times
10^7$~yr in the case of NGC~4570.  Considering the rotation period at
the ILR ($4.7 \times 10^6$~yr), as well as the very small radial
extent of the ILR ring (its diameter is $\sim 0\farcs1$), argues in
favor of the nuclear ring having formed {\it after} partial
destruction of the bar.

\section{Conclusions}

We have presented photometric evidence for the presence of two rings
in the edge-on S0 galaxy NGC~4570. The main disc of this galaxy has an
inner cut-off, inside of which a bright nuclear disc resides.  The
locations of the rings are consistent with the Inner Lindblad and
Ultra Harmonic resonances induced by a rapidly tumbling, triaxial
potential with a pattern speed of $349 \kmskpc$. The locations of the
resonances correspond to local features in the morphology, colours and
kinematics of NGC~4570. In particular, the outer disc has its inner
cut-off close to CR, consistent with the picture that this cut-off
originates from the bar (early-type bars generally end near CR, Combes
\& Elmegreen 1993).  The line strengths and colours of the central
nucleus (inside the ILR) suggests a metal-rich and rather young
stellar population. The ring at the ILR is bluer than its immediate
surrounding, which together with the complete absence of any observed
ISM, indicates that it consists of a relatively young ($\lta 2$ Gyr)
stellar population.

We have thus proposed a tentative scenario for the formation and
secular evolution of NGC~4570. We have argued that a rapidly tumbling
bar formed in NGC~4570 driving gas inwards thus forming part of the
central structure (e.g., cusp and nuclear disc). Due to this central
mass accumulation a strong ILR is created, and the bar then weakened.
This was followed by the formation of an ILR gas ring which underwent
subsequent star formation. The observed feature at the UHR also seems
to be a genuine ring feature, since both observations and N-body
simulations have shown that rings often form at the UHR as well.
However, deeper photometric observations and a comparison of observed
kinematics at the UHR with N-body simulations are required to confirm
this hypothesis.

It is intriguing that no ISM is observed in NGC~4570, whereas the
formation of rings clearly requires the presence of gas. However,
estimates of the mass of the rings, do not require a large gas supply,
but are consistent with them having formed from stellar mass loss of the
older population of stars.

Several {\it spiral} galaxies with double disc structure and/or
multiple rings whose locations are consistent with the resonant radii
of a single pattern speed are known: Emsellem \etal (1996) presented
the case of the Sombrero galaxy, M~104, an Sa which exhibits UHR and
OLR rings as well as two distinct discs.  The Sb galaxy NGC 7217
harbors three rings (at ILR, UHR, and OLR), but lacks any photometric
evidence for a bar (Athanassoula 1996).  Freeman type II discs and
stellar rings have also been observed in a significant number of {\it
  lenticular} galaxies: Seifert \& Scorza (1996) detected rings or
double disc structures in $60$\% of their sample of 15 S0s. Although
this sample has to be completed and higher spatial resolution data
obtained (e.g., to search for nuclear discs and rings), this number is
consistent with the percentage of more face-on disc galaxies which
show photometric evidence for a bar or oval distortion (Combes 1995
and references therein). All this strongly suggests that double disc
structures are common and most likely the result of bar driven secular
evolution.  In this paper we have presented the case of NGC~4570, the
first S0 galaxy known that shows the presence of {\it two} stellar
rings in addition to a double disc structure. We have discussed a
secular evolution scenario to form this complex morphology, and which
may also apply to other S0 galaxies with double disc structure and/or
ring(s).

\section*{Acknowledgments}

This paper has strongly benefited from discussions with F. Combes, M.
Sevenster, and D. Friedli. We are grateful to K. Kuijken for kindly
providing the UGD software. We thank M. Sevenster and T. de Zeeuw for
a careful reading of the manuscript, and the referee for his comments
that helped improving the paper significantly. FCvdB was supported by
the Netherlands Foundation for Astronomical Research (ASTRON), \#
782-373-055, and by a Hubble Fellowship, \#HF-01102.11-97A, awarded by
STScI.

\label{lastpage}
\end{document}